\documentclass[twocolumn,aps,floatfix,superscriptaddress,prb,showpacs]{revtex4}
\usepackage{amsmath, amsthm, amssymb, graphicx}
\input epsf.sty
\begin{document}

\def\g{\gamma}
\def\r{\rho}
\def\w{\omega}
\def\wo{\w_0}
\def\wp{\w_+}
\def\wm{\w_-}
\def\t{\tau}
\def\av#1{\langle#1\rangle}
\def\pf{P_{\rm F}}
\def\pr{P_{\rm R}}
\def\F#1{{\cal F}\left[#1\right]}

\title{Dynamic phase transition in the three-dimensional kinetic Ising model in an oscillating field}

\author{Hyunhang Park and Michel Pleimling}
\affiliation{Department of Physics, Virginia Tech, Blacksburg, Virginia 24061-0435, USA}

\date{\today}

\begin{abstract}
Using numerical simulations we investigate the properties of the dynamic phase transition that is encountered
in the three-dimensional Ising model subjected to a periodically oscillating magnetic field. The values of
the critical exponents are determined through finite-size scaling. Our results show that the studied
non-equilibrium phase transition belongs to the universality class of the equilibrium three-dimensional
Ising model.
\end{abstract}

\maketitle

\section{Introduction}
In the last fifty years our understanding of equilibrium critical phenomena has developed to a point where
well established results are available for a range of different situations. In particular, the origin of
and the difference between equilibrium universality classes is now well understood.
Far less is known, however, on the properties of non-equilibrium phase transitions taking place
in interacting many-body systems that are far from equilibrium. As for the equilibrium situation,
a continuous non-equilibrium phase transition is also characterized by a set of critical indices.
Absorbing phase transitions encountered in reaction-diffusion systems \cite{Hen08,Odo08} 
and phase transitions found in driven diffusive systems \cite{Sch95,Daq12} are well studied
examples of continuous phase transitions taking place far from equilibrium. Even though 
the values of the critical exponents have been determined in many cases, a classification of non-equilibrium
phase transitions into non-equilibrium universality classes is far from complete.

A different type of non-equilibrium criticality is encountered when kinetic ferromagnets 
are subjected to a periodically oscillating magnetic field \cite{Cha99,Ach05}. When increasing the frequency
of the field, a phase transition takes place between a dynamically disordered phase at low frequencies, where the
ferromagnet is able to follow the changes of the field, and a dynamically ordered phase at high frequencies,
where the magnetic system does not have time to adjust to the magnetic field before it changes its orientation.
In the dynamically disordered phase the magnetization oscillates around zero, but in the dynamically ordered phase
the magnetization keeps oscillating around a non-vanishing average value. This type of phase transition has
been studied in a range of systems, with applications in chemical, physical, and materials systems
as well as in ecology, and this both theoretically \cite{Ach97,Sid98,Kor00,Jan03,Ach03,Can06,Oma10}
and experimentally \cite{Jia95,Rob08}. Most of these studies looked at general properties of the
dynamical ordering, without trying to fully characterize the corresponding continuous transition through
the determination of the critical indices. It is only for the two-dimensional kinetic Ising model in 
a periodically oscillating field that the critical exponents have been measured \cite{Kor00}. Interestingly,
even though the studied phase transition is a non-equilibrium transition, the values of the exponents were
found to coincide with the values of the universality class of the equilibrium two-dimensional Ising model.
This result agrees with an earlier conjecture \cite{Gri85}, based on the symmetry of the problem, as well as
with an investigation of the time-dependent Ginzburg-Landau model with a periodically changing field \cite{Fuj01}.
From the symmetry argument given in Ref. [\onlinecite{Gri85}] one expects also for other cases that this dynamic
phase transition falls into the universality class of the equilibrium counterpart, but this has not yet been
confirmed. Interestingly, we showed in a recent study \cite{Par12} that the above mentioned symmetry argument breaks down when 
the dynamic phase transition takes place in systems with boundaries, yielding surface critical exponents
at the non-equilibrium phase transition that do not coincide with the corresponding equilibrium exponents.

In this short communication we present evidence that also for other bulk systems the dynamic phase
transition falls into the corresponding equilibrium universality class. This is done through a study
of the critical properties of the kinetic three-dimensional Ising model in a square-wave field.

In the next Section we briefly discuss the three-dimensional kinetic Ising model and the quantities
used to characterize the critical properties at the dynamic phase transition. Section III contains
the results of our numerical study. We conclude in Section IV.

\section{Model}

The Hamiltonian of the three-dimensional kinetic Ising model is given by the expression
\begin{eqnarray}
{\mathcal H}=-J\sum_{\langle\textbf{x},\textbf{y}\rangle}S_{\textbf{x}}S_{\textbf{y}}-H(t)\sum_{
\textbf{x}}S_{\textbf{x}}
\end{eqnarray}
where $S_{\textbf{x}}=\pm1$ is the usual Ising spin located at site $\textbf{x}$. The
first sum in that expression runs over bonds connecting nearest neighbor spins. $J>0$
is a ferromagnetic coupling and $H(t)$ is a spatially uniform periodically oscillating
magnetic field. Following [\onlinecite{Kor00}] we use a square-wave field of
amplitude $H_0$ and half-period $t_{1/2}$. For the data presented in this paper
we choose temperature and magnetic field strength so that the system is in the multidroplet regime:
$T=0.8~T_c$ and $H_0=0.4~J$, with the critical temperature
$T_c = 4.5115 J/k_B$ ($k_B$ is the Boltzmann constant).

\begin{figure} [h]
\includegraphics[width=0.90\columnwidth]{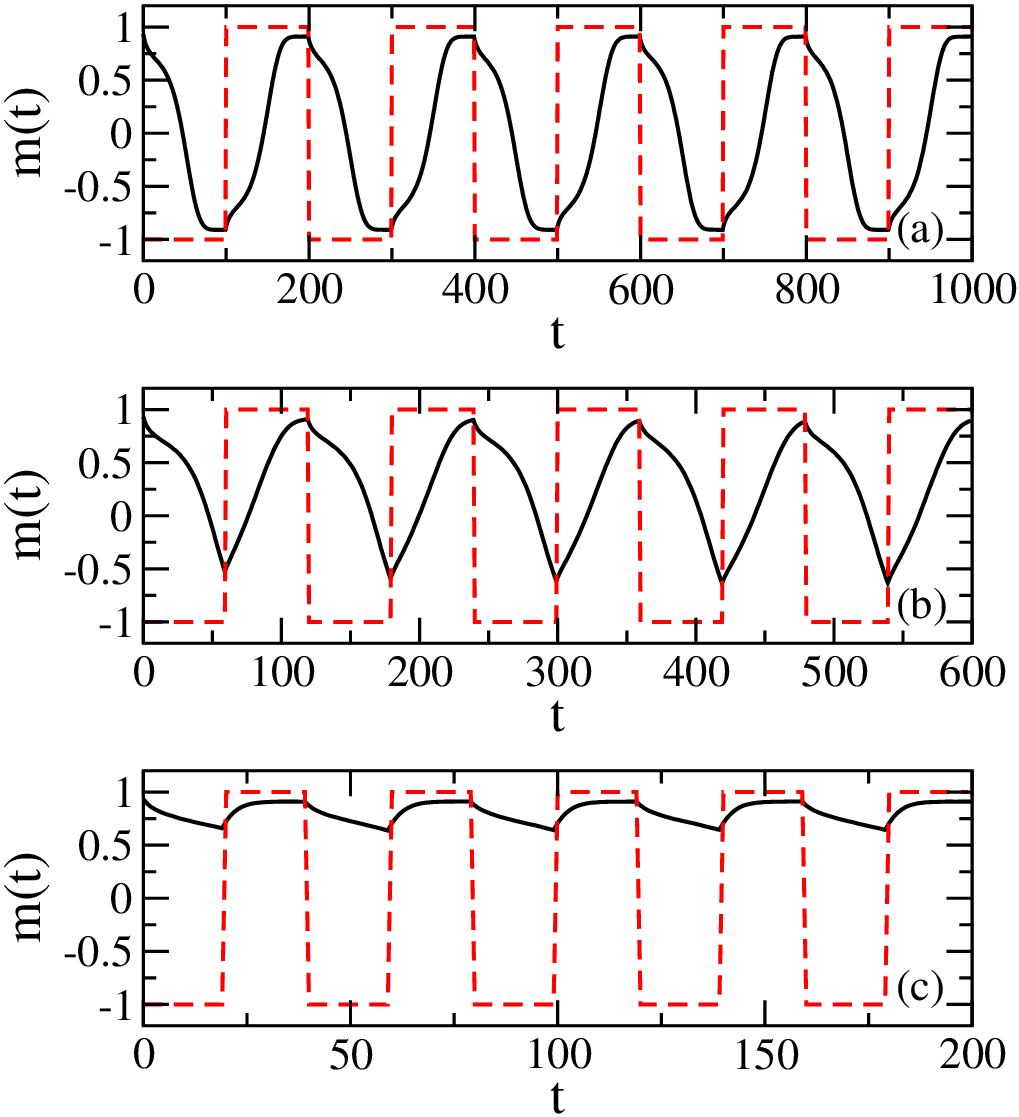}
\caption{\label{fig1} (Color online)
Time-dependent magnetization of the three-dimensional kinetic Ising model in a square-wave field of
half-period (a) $t_{1/2} = 100$, (b) $t_{1/2} = 60$, and (c) $t_{1/2} = 20$. Starting with the fully magnetized sample,
the first five periods are shown for each case. The red dashed line is $H(t)/H_0$ where $H_0$ is the amplitude
of the magnetic field. Whereas for case (a) we are in the dynamically ordered phase, for case (c) the system
is dynamically disordered. Case (b) is close to the dynamical phase transition.
The parameters are $T=0.8~T_c$ and $H_0=0.4~J$, the system contains $96^3$ spins.
The unit of time is one Monte Carlo Step.
}
\end{figure}

Let us pause briefly in order to summarize the mechanism underlying the dynamical ordering
observed in the kinetic ferromagnets, see Fig. \ref{fig1}. 
Assuming the magnetization being aligned with the external field,
a reversing of the direction of the field results in a metastable state from which the system
tries to escape by nucleating droplets that are pointing in the same direction as the field.
Whereas in the dynamically disordered phase, the ferromagnet is able to reverse its magnetization before the
field is changed again, see Fig. \ref{fig1}a, in the dynamically ordered phase the system is still in the metastable state when
the field direction switches back, thereby yielding a time dependent magnetization that oscillates
around a finite value, as shown in Fig. \ref{fig1}c. This competition between the magnetic field and the metastable state is captured
by the variable
\begin{eqnarray}
\Theta=\frac{t_{1/2}}{\langle\tau\rangle}
\end{eqnarray}
where $\langle\tau\rangle$ is the 
metastable lifetime. For the kinetic Ising model $\Theta$ plays the same role as that played by temperature
in the equilibrium system. Changing $t_{1/2}$ changes the value of $\Theta$. Consequently, there
is a critical value $\Theta_c$ where the dynamic phase transition between the dynamically ordered
and the dynamically disordered phases takes place, see Fig. \ref{fig1}b.

\begin{figure} [h]
\includegraphics[width=0.90\columnwidth]{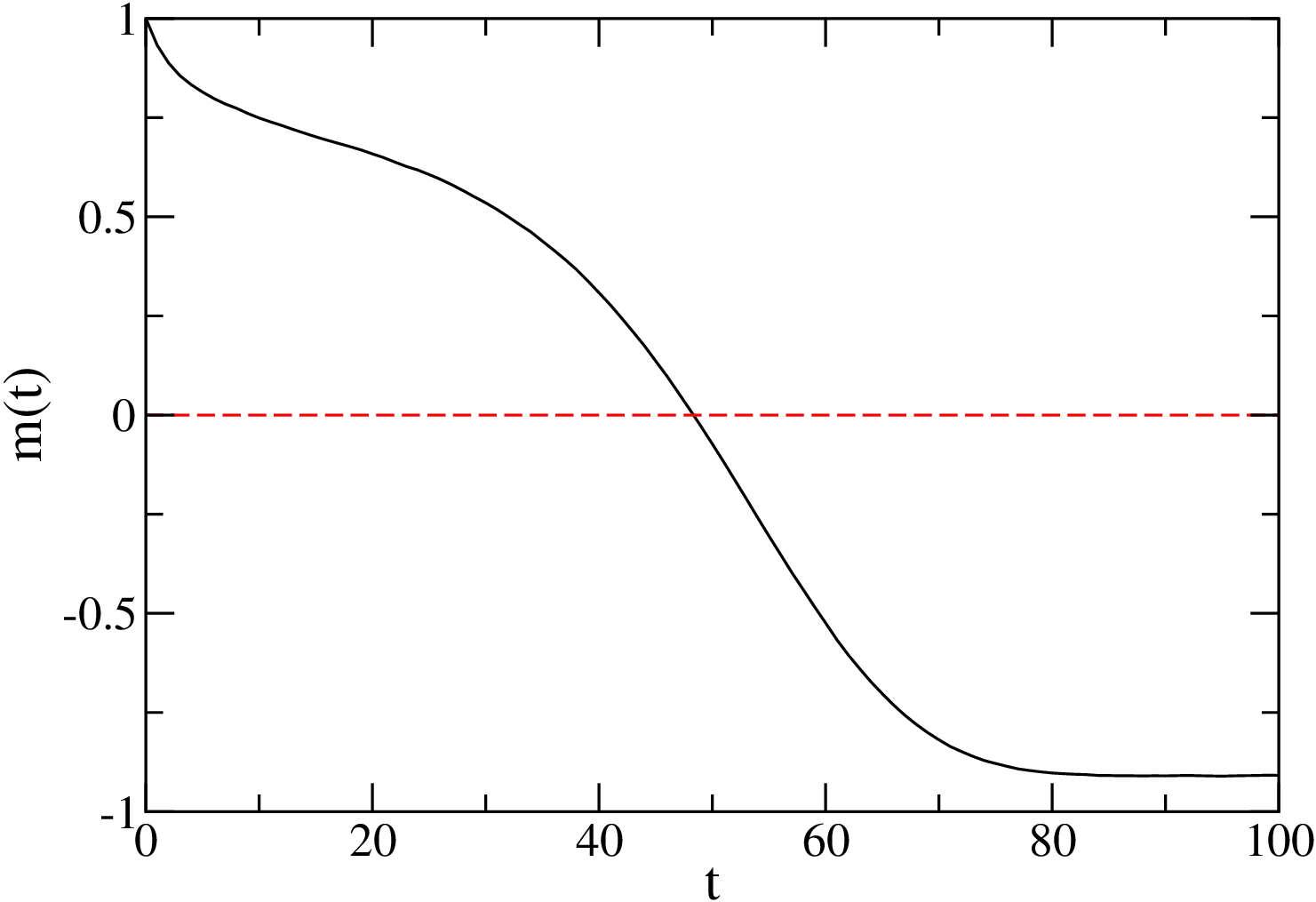}
\caption{\label{fig2} (Color online)
Determination of the metastable lifetime. After preparing the system in a fully ordered state a constant
magnetic field is applied that reverses the magnetization. The metastable lifetime is
defined as the first-passage time to zero magnetization. The parameters are $T=0.8~T_c$ and $H_0=0.4~J$,
the data shown have been obtained for a system composed of $96^3$ sites. The metastable lifetime 
$\langle\tau\rangle = 47.05$ is independent of the system size.
The unit of time is one Monte Carlo Step.
}
\end{figure}

In order to determine the metastable lifetime we consider an equilibrated system in a field of constant
strength $H_0$. Reversing the sign of the field renders the system metastable. The metastable lifetime 
is then obtained as the average first-passage time to zero magnetization, see Fig. \ref{fig2}. For our system parameters,
$T=0.8~T_c$ and $H_0=0.4~J$, we obtain that $\langle\tau\rangle = 47.05$, where the time is measured
in Monte Carlo Step, with one Step corresponding to updating on average every spin once.

In order to elucidate the critical properties of the dynamic phase transition we study a range of different
quantities. The main quantity of interest is the period-averaged magnetization
\begin{eqnarray}
Q=\frac{1}{2t_{1/2}}\oint m(t)dt
\end{eqnarray}
where the integration is performed over one period of the oscillating field.
The time dependent magnetization is given by
\begin{eqnarray}
m(t)=\frac{1}{N}\sum_{\textbf{x}}S_{\textbf{x}}(t)~.
\end{eqnarray}
We thereby consider cubic systems with periodic boundary conditions and linear extend $L$, so that $N = L^3$.
The order parameter $\langle | Q | \rangle$ is then obtained after averaging both over time (i.e. an average over many periods) 
and over different realizations of the noise.

Some quantities of interest are directly obtained from the order parameter. Thus we calculate the 
Binder cumulant
\begin{eqnarray}
U=1-\frac{\langle Q^4\rangle}{3\langle Q^2\rangle^2}
\end{eqnarray}
that we use to determine the critical point, see Section III. The susceptibility describing the fluctuations
of the order parameter is given by
\begin{eqnarray}
\chi^Q=N(\langle Q^2\rangle-\langle|Q|\rangle^2)~.
\end{eqnarray}
We also calculated the period-averaged energy
\begin{eqnarray}
E = - \frac{1}{2t_{1/2}}\oint \frac{J}{N} \sum_{\langle \textbf{x},\textbf{y}\rangle} S_{\textbf{x}}S_{\textbf{y}} ~dt
\end{eqnarray}
and its fluctuations
\begin{eqnarray}
\chi^E=N(\langle E^2\rangle-\langle E \rangle^2)~.
\end{eqnarray}

Close to the critical point, the order parameter and the response functions $\chi^Q$ and $\chi^E$ display an
algebraic behavior:
\begin{eqnarray}
\langle | Q | \rangle & \sim & (\Theta_c - \Theta )^\beta \\
\chi^Q & \sim & \left| \Theta_c - \Theta \right|^{-\gamma} \\
\chi^E & \sim & \left|\Theta_c - \Theta \right|^{-\alpha}
\end{eqnarray}
with the critical exponents $\beta$, $\gamma$, and $\alpha$.

\section{Results}

\begin{figure} [h]
\includegraphics[width=0.90\columnwidth]{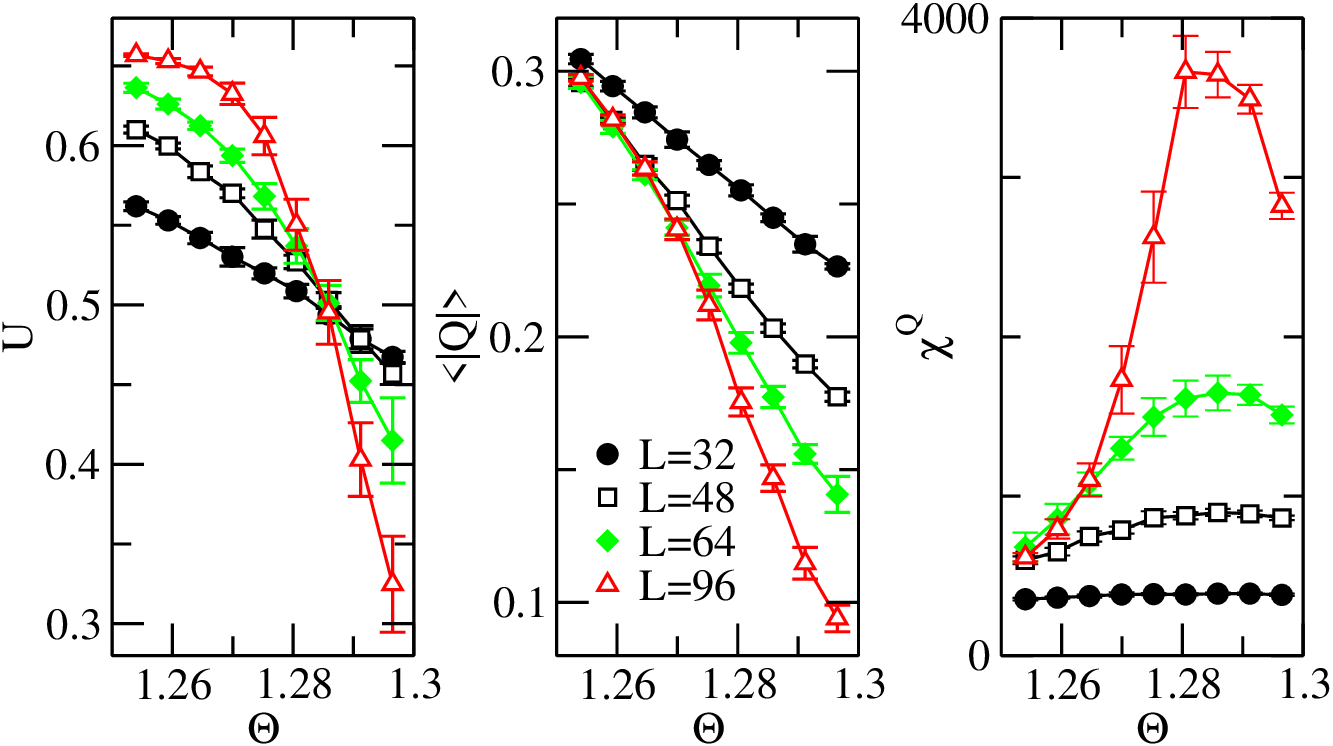}
\caption{\label{fig3} (Color online)
(a) Binder cumulant, (b) order parameter, and (c) susceptibility as a function of $\Theta$ for systems
with different linear extend $L$. The cumulants for the different system sizes cross at $\Theta_c 
= 1.285$. In addition, the susceptibility displays in the finite systems a maximum close to that
value of $\Theta$. The data result from averaging over typically 100000 periods. 
}
\end{figure}

As a necessary prerequisite for a reliable determination of the critical exponents 
we need to know the location of the critical point with high precision.
As usual for continuous phase transitions the various quantities show a characteristic
behavior in finite systems when approaching the critical point. This is illustrated in Fig. \ref{fig3} for
the Binder cumulant, the order parameter, as well as for the fluctuations of the order
parameter. Even though one can get an estimate for the critical value $\Theta_c$ by analyzing
the shift of the maximum of the susceptibility when changing the system size, the most reliable
way to determine the critical point is to study the Binder cumulant. As seen in Fig. \ref{fig3}a, for the kinetic
three-dimensional Ising model the data sets obtained for different system sizes all cross at a
common value $\Theta = 1.285$. This value being compatible with the finite-size shifts of the
position of the maxima in $\chi^Q$ and $\chi^E$, we take this value as the critical value $\Theta_c$
where the dynamic phase transition takes place.

In order to determine the values of the critical exponents, we perform a finite-size scaling analysis \cite{Bin90},
similar to what has been done previously for the two-dimensional kinetic Ising model \cite{Kor00}.
Close to the critical point systems of different sizes should display the following scaling
behavior:
\begin{eqnarray}
\langle|Q|\rangle&=&L^{-\beta/\nu}{\mathcal{F}}_{\pm}(\theta L^{1/\nu})\\
\chi^Q&=&L^{\gamma/\nu}{\mathcal{G}}_{\pm}(\theta L^{1/\nu})\\
\chi^E&=&L^{\alpha/\nu}{\mathcal{I}}_{\pm}(\theta L^{1/\nu})
\end{eqnarray}
where $\theta=\frac{|\Theta-\Theta_c|}{\Theta_c}$ is the reduced control parameter and $\nu$
is the critical exponent describing the divergence of the correlation length when approaching
the critical point.
${\mathcal{F}}_{\pm}$, ${\mathcal{G}}_{\pm}$, and ${\mathcal{I}}_{\pm}$ are scaling functions, where the
sign $\pm$ correspond to $\Theta\gtrless\Theta_c$. Comparison of data obtained 
at the critical point $\Theta = \Theta_c$ should yield for all three quantities 
an algebraic dependence on the system size, which allows to directly determine the critical exponents
from the slopes in a double logarithmic plot.

\begin{figure} [h]
\includegraphics[width=0.90\columnwidth]{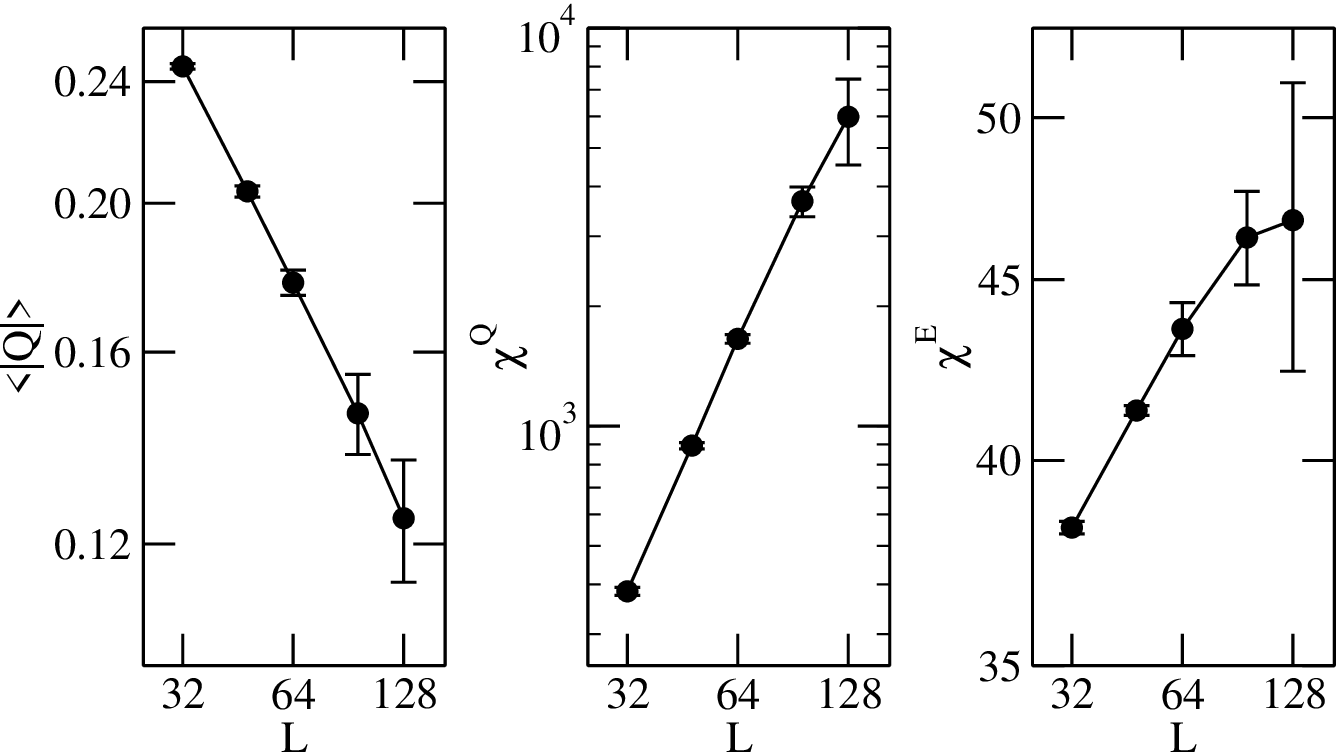}
\caption{\label{fig4} 
(a) Order parameter $\langle|Q|\rangle$, (b) order parameter susceptibility $\chi^Q$, and (c) energy fluctuations $\chi^E$
at the critical point $\Theta = \Theta_c$
as a function of system size. The slopes in a double logarithmic plot give the values of the
critical exponents $\beta/\nu$, $\gamma/\nu$, and $\alpha/\nu$. The data have been obtained by averaging
over at least 200000 periods. 
System sizes are measured in multiples of the lattice constant.
}
\end{figure}

As shown in Fig. \ref{fig4} all three quantities indeed depend algebraically on the system size. 
From the slopes we obtain the values $\beta/\nu = 0.51(2)$, $\gamma/\nu = 1.96(4)$, and $\alpha/\nu = 0.18(3)$
(for the specific heat we did not take into account the largest system size, due to the rather poor quality
of the data).
Comparing these values with the well known equilibrium values for the three-dimensional Ising model
$\beta = 0.33$, $\gamma = 1.24$, $\alpha = 0.108$, and $\nu = 0.63$, we see that the set of exponents
we obtain for the dynamic phase transition of the kinetic Ising model indeed agree with the equilibrium
values $\beta/\nu = 0.52$, $\gamma/\nu = 1.97$, and $\alpha/\nu = 0.17$.
This confirms that for the problem at hand the symmetry argument given in Ref. [\onlinecite{Gri85}]
is also valid in three space dimensions.

\section{Conclusion}

Classifying and understanding universality classes of non-equilibrium phase transitions remains an
important task. For some types of systems, as for example absorbing phase transitions, 
field theoretical methods can be used successfully, which then allow not only to calculate
critical quantities but also to understand what features of a given system are universal. Still,
the classification of non-equilibrium universality classes is far from complete, and much work remains
to be done before a more complete understanding of the origin of universality far from equilibrium
is achieved.

The dynamic phase transition encountered in magnetic systems in an oscillating field constitutes an interesting case.
Indeed, it was shown that for the two-dimensional kinetic Ising model
the critical exponents at this non-equilibrium phase transition coincide with those of the equilibrium
Ising model \cite{Kor00}. This result is supported by a symmetry argument \cite{Gri85}. On the other
hand, the presence of surfaces results in non-equilibrium surface universality classes that differ
from those encountered in the equilibrium semi-infinite Ising model \cite{Par12}.

In this work we have shown that also for the dynamic phase transition in the three-dimensional 
kinetic Ising model the same critical exponents as for the three-dimensional equilibrium model are obtained.
This lends further support to the symmetry argument given in [\onlinecite{Gri85}]. 
It also stresses the importance of symmetries in establishing universalities in non-equilibrium
critical phenomena.

\begin{acknowledgments}

We thank Per Arne Rikvold for suggesting this publication as well as for helpful discussions.
This work is supported by the US National
Science Foundation through grants DMR-0904999 and DMR-1205309.

\end{acknowledgments}

\end{document}